\documentclass[12pt]{article}
\parskip=\medskipamount

\usepackage{hyperref}
\usepackage{amssymb}
\usepackage{graphicx}
\def \N {{\cal N }}

\def \un{\underline}

\newcommand {\cB}{{\cal B}}

\newcommand {\cD}{{\cal D}}

\newcommand {\cF}{{\cal F}}

\newcommand {\cL}{{\cal L}}
\newcommand {\cM}{{\cal M}}
\newcommand {\cN}{{\cal N}}

\newcommand {\cP}{{\cal P}}

\newcommand {\cW}{{\cal W}}

\newcommand{\bW}{{\bf W}}

\def\a{\alpha}
\def \bi{\bibitem}
\def \ci{\cite}
\def\b{\beta}

\def\d{\delta}

\def\f{\phi}

\def\G{\Gamma}

\def\m{\mu}
\def\n{\nu}
\def\o{\omega}
\def\p{\pi}
\def\q{\theta}
\def\r{\rho}
\def\s{\sigma}

\def\x{\xi}
\def\z{\zeta}

\def\F{\Phi}
\def\J{\Psi}

\def\P{\Pi}

\def\S{\Sigma}

\newcommand{\ad}{{\dot{\alpha}}}                           
\newcommand{\bd}{{\dot{\beta}}}                            
\newcommand{\ve}{\varepsilon}                            

\newcommand{\hf}{\frac12}

\newcommand{\vf}{\varphi}
\newcommand{\sect}[1]{\setcounter{equation}{0}\section{#1}}

\newcommand{\be}{\begin{equation}}
\newcommand{\ee}{\end{equation}}
\newcommand{\bea}{\begin{eqnarray}}
\newcommand{\eea}{\end{eqnarray}}
\newcommand{\non}{\nonumber}

\newcommand{\bm}[1]{\mbox{\boldmath$#1$}}
\def \foot {\footnote}

\begin{document}

\begin{titlepage}

\begin{flushright}
hep-th/0508098\\
August, 2005
\end{flushright}
\vspace{5mm}

\begin{center}
{\Large\bf  
Effective action of $\bm \b$-deformed $\bm {\cN = 4}$ SYM
theory
\\ \vspace{2mm}
and AdS/CFT
}
\end{center}

\begin{center}

{\large 
 S. M. 
Kuzenko${}^{a,}$\footnote{kuzenko@cyllene.uwa.edu.au}
and A. A. Tseytlin${}^{b,}$\footnote{Also at Imperial College, London and 
Lebedev Institute, Moscow. tseytlin@mps.ohio-state.edu}
}
\vspace{2mm}

${}^a$\footnotesize{
{\it School of Physics M013, The University of Western Australia,\\
35 Stirling Highway, Crawley W.A. 6009, Australia}} \\
${}^b$\footnotesize{
{\it Department of Physics, The Ohio State University,\\
Columbus, OH 43210, USA}}
\vspace{2mm}

\end{center}
\vspace{5mm}

\begin{abstract}
\baselineskip=14pt
We compute  the one-loop  
effective action in ${\cN = 1}$ superconformal $SU(N)$ 
gauge  theory 
which is an exactly marginal deformation   of
 the ${\cN = 4}$ SYM. We consider an 
  abelian background of constant 
${\cN = 1}$  gauge field  and single chiral  scalar. 
While  for  finite $N$  the effective action depends 
non-trivially on the deformation parameter $q= e^{i \pi \b}$, 
this dependence disappears  in the large $N$ limit
if the  parameter $\b$ is real.
This conclusion matches  the strong-coupling prediction 
coming from the  form of   a  D3-brane probe action in the dual 
supergravity background:  for 
the simplest  choice of the 
D3-brane position  the probe  action  happens to be 
 the same as  for 
a   D3-brane in $AdS_5 \times S^5$ placed  
parallel to the boundary of  $AdS_5$.  
This  suggests that in the real $\b$ deformation case
 there  exists   a  large $N$ 
non-renormalization theorem for the $F^4$ term in the action.

\end{abstract}

\vfill
\end{titlepage}

\newpage
\setcounter{page}{1}
\renewcommand{\thefootnote}{\arabic{footnote}}
\setcounter{footnote}{0}
\def \P {\Phi}

\sect{Introduction}

The  study of AdS/CFT duality  for less supersymmetric cases
was recently boosted  by the discovery of the 
supergravity background \ci{LM} dual  to the exactly marginal
$\cN=1$ superconformal $\b$-deformation  \ci{LS} 
of the maximally supersymmetric $SU(N)$  SYM theory
(earlier work on this gauge theory and its 
supergravity dual
appeared  in  \ci{old}  and \ci{ahar,NIPR}). 
The most immediate  implication   of the large $N$  AdS/CFT duality is the matching 
between  the anomalous  dimensions of  single-trace  composite
operators  and the corresponding spectrum of string energies.
This matching was demonstrated  
 in a certain  ``semiclassical'' limit  in 
\ci{FRT1,mat} (using, in particular, 
   gauge-theory  results of \ci{RR}).

Some     properties of  anomalous dimensions and
 correlation functions  of the 
$\b$-deformed  gauge theory  were recently studied    in 
\ci{FG,PSZ,RSS,MPS}.
Here we complement this work by considering 
 the one-loop low-energy effective action for a  simple
 gauge field $F$ and scalar $\P$  background 
 on a Coulomb branch \ci{BE,DH}  of  the $\b$-deformed 
 theory.\foot{Second-derivative term in the low-energy effective
  action  at a generic point of 
  the Coulomb branch was considered  in \ci{DH}. Here we will be interested in 
   4-derivative $F^4$, etc.,  terms.} 
 This allows us to  compare the $F^4/|\P|^4$ term in this 
 action  with the corresponding term in the 
 action of a D3-brane probe
 placed in the deformed $(AdS_5 \times S^5)_\b$ 
 background.\foot{Matching  of constant scalar potential 
  term in D3-brane probe action 
 with 1-loop  correction in 
 the general deformed gauge theory away from superconformal point 
  was observed  earlier 
  in \ci{ahar}.}
 As is well known, in the case of undeformed  $\cN=4$ SYM theory 
 the two terms agree \ci{mald} 
  (for a review and extensions  see \ci{BKT}), 
  and this may be interpreted as a
 manifestation of a non-renormalization theorem \ci{DS}.

 For the simplest  abelian gauge theory background we shall consider 
 below only one of the three chiral scalar fields 
  will  be chosen to have 
 a non-zero   value. In the dual supergravity picture 
 this translates into the position of the D3-brane probe  
 on $(S^5)_\b$ being 
 at $\mu_1=1, \ \mu_2=\mu_3=0$ (in the notation of \ci{LM}) 
 with the three other isometric angles
 being trivial. In this case the inspection of the deformed
 background in \ci{LM}
  shows that all the dependence on the (in general,
 complex) deformation parameter $\b$  drops out, i.e. the D3-brane
 probe action happens to be   the same as in the 
   $AdS_5 \times  S^5$ case.

 The coefficient of the $F^4/|\P|^4$ term in the 
 one-loop  $SU(N)$ gauge theory effective action we 
 shall compute below 
 has, in general, a non-trivial dependence on $\b$.
 However, this  dependence completely 
  drops out in the large $N$ limit, provided $\b$  is real. 
  In fact, the  large 
  $N$ limit of the 1-loop effective action in the real deformation case 
turns  out to be the same  as in the undeformed 
 SYM theory.  This agrees with  the strong-coupling  prediction 
 coming from the D3-brane probe action, with a 
  plausible explanation 
 of this  matching  being  the existence of a 
 non-trivial large $N$  non-renormalization theorem.

 The case of complex $\beta$  is different: here  the dependence 
 of the one-loop gauge theory effective action on the deformation parameter survives 
 the large $N$ limit and thus disagrees  with  the form of  the D3-brane  action. 
 This provides  another indication that   the complex $\beta$ deformation  case 
 is  more complicated  than the real $\beta$ one, and that the implications of the 
 AdS/CFT  duality here  are  much harder to  uncover
 (other complications of complex $\beta$ case 
 are  lack of integrability on both gauge  theory and string theory sides, 
 need to use S-duality \cite{LM}  to construct 
 the  string background implying  lack of useful  perturbative 
 definition of the corresponding string theory, etc., \cite{FRT1}).

Below in section 2 we shall review the structure of the 
$\b$-deformed   gauge theory and write down the general 
expression for its 1-loop effective action in an abelian background. 
In section 3 we shall find the   explicit form of the term quartic in the
gauge field strength and analyze its dependence on the deformation 
parameter $q= e^{i \pi \beta}$  and $N$.  
Appendices A, B and C 
contain some technical details while in Appendix D we present the form 
of the second-derivative term in the effective action  in the case of a 
more general diagonal 
abelian background.

\sect{One-loop effective action in the $\beta$-deformed $\cN$ = 4 SYM theory}
${}$Following   the $\cN=1$ superspace conventions of \cite{BK,KM2} 
the $\b$-deformed $\cN=4$ 
$SU(N)$ SYM theory 
is described by the action 
\bea
S &=& 
\int {\rm d}^8 z \, {\rm Tr}\,( \F_i{}^\dagger \,\F_i )
+   \frac{1}{g^2} \int {\rm d}^6 z \, 
{\rm Tr} (
\cW^\a \cW_\a ) \non \\
&+& \left\{
h  \int {\rm d}^6 z \, {\rm Tr} (
q\, \F_1 \F_2 \F_3 
-q^{-1}\,
\F_1 \F_3 \F_2 )
~+~ {\rm c.c.} \right\}~, \ \ \ \ \  q \equiv 
{\rm e}^{{\rm i} \p \b }\ ,
\label{b-deformed-N=4SYM}
\eea
where 
$q $ is the deformation parameter, $g$ is the gauge coupling constant, 
and $h$ is related to $g$ and $q$  by the conformal invariance
condition ($h=g$ in the undeformed theory when  $q=1$).  
Here $\F_i = \F^\m_i (z) T_\m$  \ ($i=1,2,3$) 
are the  {\it covariantly chiral}
superfields, ${\bar \cD}_\ad \F_i=0$.\footnote{The $SU(N)$
generators $T_\m =(T_\m)^\dagger$ are normalized so that 
${\rm Tr} \,(T_\m\, T_\n) = \d_{\m \n}$.}  
The covariantly chiral field strength $\cW_\a$, 
${\bar \cD}_\ad \cW_\a =0$, 
is associated with  the gauge  covariant derivatives
\be
\cD_A = (\cD_a, \cD_\a , {\bar \cD}^\ad ) 
= D_A +{\rm i}\, \G_A~, \qquad 
\G_A = \G^\m_A (z) T_\m~,
\label{gcd}
\ee
where $D_A$ are the flat covariant 
derivatives.
The  gauge covariant derivatives satisfy the following algebra:
\bea
& \{ \cD_\a , \cD_\b \} 
= \{ {\bar \cD}_\ad , {\bar \cD}_\bd \} =0~, \qquad 
\{ \cD_\a , {\bar \cD}_\bd \} = - 2{\rm i} \, \cD_{\a \bd}~, \non \\
& [ \cD_\a , \cD_{\b \bd}] = 2 {\rm i} \ve_{\a \b}\,{\bar \cW}_\bd ~, 
\qquad 
[{\bar \cD}_\ad , \cD_{\b \bd}] = 2{\rm i} \ve_{\ad \bd}\,\cW_\b ~ , 
\non \\
& [ \cD_{\a \ad}, \cD_{\b \bd} ] 
={\rm i}\, \cF_{\a \ad, \b \bd} 
= - \ve_{\a \b}\, {\bar \cD}_\ad {\bar \cW}_\bd 
-\ve_{\ad \bd} \,\cD_\a \cW_\b~. 
\label{N=1cov-der-al}
\eea
The spinor field strengths $\cW_\a$ and ${\bar \cW}_\ad$ 
obey the Bianchi  identities
$\cD^\a \cW_\a = {\bar \cD}_\ad {\bar \cW}^\ad$.

The extrema of the scalar potential (the Coulomb branch) 
are  described by the equations 
(here $\F_i$ are the first components of the chiral superfields)
\be
 \sum_i [ \F_i  \, ,\F_i{}^\dagger  ] =0 ~, 
\qquad 
q \, \F_i \F_{i+1} -q^{-1} \,\F_{i+1} \F_i 
= {1\over N}  \, (q - q^{-1} ) \,{\rm Tr} (\F_i \F_{i+1} )
\, {\bf 1} ~.
\ee
In what follows, 
we shall consider the simplest 
special  solution 
 \be \F_1 \equiv \F \ , \ \ \ \ \ \ \ 
   \F_2=\F_3=0 \ , \ee 
where $\F$ is a  diagonal traceless $N \times N$ matrix.
${}$For such special background one  is able
to study only a limited class of gauge-invariant quantities like effective
action and  anomalous dimensions of certain scalar operators.

To quantize the theory, we 
use the $\cN=1$ 
background field formulation \cite{GGRS} and 
split the dynamical variables into the background 
and quantum ones (for a summary and the gauge
conditions chosen 
see Appendix A)  
\bea
  \F_i ~ \to ~ \F_i +\vf_i ~, 
\qquad   \quad
\cD_\a ~ \to ~ {\rm e}^{-g\,v} \, \cD_\a \, {\rm e}^{g\, v}~, 
\qquad
{\bar \cD}_\ad ~ \to ~ {\bar \cD}_\ad~,
\label{bq-splitting}
\eea
with lower-case letters used for 
the quantum superfields. 
Choosing  $\F_2=\F_3=0$ and $\F_1 \equiv \F \neq 0$,
the  quadratic part of the gauge-fixed action is 
$S + S_{\rm gf}$ is 
\bea 
S^{(2)} + S_{\rm gf} 
 &=&  - \frac{1}{2}
\int {\rm d}^8 z \,{\rm Tr}\,
\Big( v \,\Box_{\rm v}  v - g^2 \,v\,[\F^\dagger, [\F, v]] \Big) 
\non \\
&&+ \int {\rm d}^8 z \,
{\rm Tr}\,\Big( \vf_1^\dagger \,\vf_1 
- g^2 \,[\F^\dagger , [\F, \vf_1^\dagger ]] \, 
(\Box_+)^{-1}\, \vf_1 \Big) 
~+~ \dots  
\label{quad-prel}
\\
&&+
 \int {\rm d}^8 z \,{\rm Tr}\,
 \Big( \vf_2^\dagger \, \vf_2 
+  \vf_3^\dagger \,  \vf_3   \Big) 
+  
\left\{ 
h  \int {\rm d}^6 z \, {\rm Tr}\, \Big(
q\, \F \vf_2 \vf_3 
- {1\over q} \,\F \vf_3 \vf_2 \Big)
+ {\rm c.c.} \right\}~, \non 
\eea 
where the dots stand for the terms with derivatives
of the background (anti)chiral superfields 
 $\F^\dagger$ and $\F$.
The vector d'Alembertian,
$\Box_{\rm v}$,  is defined by
\bea 
{\Box}_{\rm v} 
&=& \cD^a \cD_a - \cW^\a \cD_\a +{\bar \cW}_\ad {\bar \cD}^\ad ~.
\label{vector-box} 
\eea

We shall choose  
the background superfields  
(i) to be  covariantly constant and on-shell, and 
(ii) to  take their values in the Cartan subalgebra of $SU(N)$.
In particular, they will satisfy the conditions 
($\cD_{(\a} \cW_{\b)} \neq 0$):
\be
[\F , \F^\dagger ] = 0~, \qquad \cD_\a \F =0~,
\qquad  \cD_a \cW_\b =0~, \qquad 
\cD^\a \cW_\a = 0~. 
\label{back-con-1}
\ee
Let us 
introduce the mass operator $\cM_{(h,q)}$ defined by
its action on a superfield $\S = \S^\m T_\m$ 
\bea
\cM_{(h,q)} \S &=& h \,(q \,\F\, \S 
- {1\over q} \,\S \,\F) -  {h\over N}  \,(q  - {1\over q} )\,
{\rm Tr} ( \F\, \S ) \,{\bf 1}~,
\label{massop} \\
\left[ \cM_{(h,q)} \, , \cM^\dagger_{(h,q)} \right]\,  \S
&=& { h {\bar h} \over N} \, (q  - {1\over q} )\,
({\bar q}  - {1\over \bar q} )
\Big\{ \F^\dagger \,{\rm Tr} ( \F\, \S ) 
-\F \, {\rm Tr} ( \F^\dagger \, \S ) \Big\}~.
\non
\eea
The commutator $[ \cM_{(h,q)} \, , \cM^\dagger_{(h,q)} ]$
does not vanish, for $q\neq \pm 1$, only when acting on 
special vectors in the Cartan subalgebra.
Then  (\ref{quad-prel}) becomes
\bea 
S^{(2)} + S_{\rm gf} 
 &=&  
\int {\rm d}^8 z \,{\rm Tr}\,\Big[
 \vf_1^\dagger \, (\Box_+)^{-1}\,
(\Box_+ -  |\cM_{(g,1)} |^2)\, \vf_1
-\hf v \, ( \Box_{\rm v}   - |\cM_{(g,1)} |^2)\, v
 \Big] 
\non \\
&+&
 \int {\rm d}^8 z \,{\rm Tr}\,
 \Big( \vf_2^\dagger \, \vf_2 
+  \vf_3^\dagger \,  \vf_3   \Big) 
+  \left\{ 
 \int {\rm d}^6 z \, {\rm Tr}\, \Big(
 \vf_3\, \cM_{(h,q)} \, \vf_2
\Big)
+ {\rm c.c.} \right\}~.
\label{quad-prel2}
\eea 
Similarly, the quadratic part of the Faddeev-Popov
ghost action takes the form
\bea 
S^{(2)}_{\rm gh} 
 &=& 
 \int {\rm d}^8 z \,{\rm Tr}
\Big[
c^\dagger 
(\Box_+)^{-1} 
(\Box_+ - |\cM_{(g,1)}|^2 )\, \tilde{c}
- \tilde{c}^\dagger 
(\Box_+)^{-1}
(\Box_+ - |\cM_{(g,1)}|^2 ) \,c \Big]~.
\label{2-gh-ac}
\eea
One should also take into account 
the Nielsen-Kallosh ghost action 
(\ref{third-ghost}).

The one-loop effective action can then be shown to be 
\bea
\G_{{\rm 1-loop}} &=& 
{ {\rm i} \over 2} \,{\bf Tr} \, 
\ln  \, ( \Box_{\rm v}   - |\cM_{(g,1)} |^2) \non \\
&+& {\rm i} \,{\bf Tr}_+ 
\ln  \, (\Box_+ - |\cM_{(h,q)} |^2 )
- {\rm i} \,{\bf Tr}_+  \ln  \,   (\Box_+ -  |\cM_{(g,1)} |^2)~.
\label{ea}
\eea
In the case of $\cN=4$ SYM, we have $h=g $ and $q=1$, and 
then only the contribution in the first line survives.
As follows from the discussion  in section 3
and Appendix C,
$\G_{{\rm 1-loop}} $ is finite. 

\sect{Evaluation of the effective action}
 
More specifically, we shall choose 
the  background scalar and vector superfields 
as 
\be
\F = g^{-1} \f \, H_0~, \ \ \ \ \  \qquad \cW_\a = W_\a \, H_0~,
\label{actual-background}
\ee
where $\f$  and $  W_\a $ are  singlet fields and 
 $H_0$ is a special generator in the Cartan subalgebra  of $SU(N)$.
The  characteristic feature of this field configuration 
is that it leaves
the subgroup $U(1) \times SU(N-1) \subset SU(N)$ 
unbroken, where $U(1)$ is associated with $H_0$
and  $SU(N-1)$ is  generated by 
$\{ H_{\un{I}}, \, E_{ \un{i} \un{j}} \}$
(see Appendix B for the notation and 
explicit form of the Cartan-Weyl basis). 

Not all components $u^\m$ of a  quantum superfield
$u$ of the form (\ref{generic}) couple to the background vector
multiplet. As follows from the identity 
\be
[H_0 , E_{ij}] ~=~ 
e\,  
\Big( 
\d_{0i}\, E_{0j} - \d_{0j}\, E_{i0} \Big)~,  \ee
\be 
e \equiv   \sqrt{N \over N-1}~,
\ee
there are $(N-1)$ superfields 
$u^{0\, \un{i} }$ of charge $+e$, and 
$(N-1)$  superfields $u^{\un{i} \, 0}$ of charge $-e$.
The rest of components of $u^\m$ are neutral.
The mass operator (\ref{massop}) 
acts on the generators  associated with the charged components as
\bea
\cM_{(h,q)} \, E_{0\un{i}} = e \f\, g^{-1} h \, \Big[
 q&-& {1\over N} \, (q  - {1\over q} )\Big] \, E_{0\un{i}}
\equiv \m^+_{(h,q)} \, E_{0\un{i}} ~, 
\non \\ 
\cM_{(h,q)} \, E_{\un{i} 0} = -e \f\, g^{-1}h\,  \Big[
{1\over  q}& + & {1\over N} \, (q  - {1\over q} )\Big] \, E_{\un{i}0}
\equiv -\m^-_{(h,q)} \, E_{\un{i}0}~.
\label{deformed-masses}
\eea
The two eigenvalues in (\ref{deformed-masses})
have the same norm if 
$|q|= 1$. 

Let us  now recall
the condition that guarantees  the one-loop anomalous dimension matrix
 for chiral  superfields
vanishes, which is the same as the UV finiteness condition\footnote{In the case
of a nonabelian background $\F$, one can obtain a general 
 expression
for the one-loop K\"ahler potential $K_1 (\F , \bar \F )$ using, for instance,
the techniques developed in the first reference in  \cite{BBP}. One can then 
explicitly evaluate $K_1 (\F , \bar \F )$   when $\F$ is diagonal matrix, e.g., 
of the form  (\ref{actual-background}). One finds indeed that the 
 K\"ahler potential is free of divergences
if eq. (\ref{finiten1}) holds. In the large $N$ limit and in the {\it real} $\b$  case,
it can be shown  that the one-loop correction to the K\"ahler potential $K_1(\F , \bar \F )$ 
 is subleading
as compared to  the 4-derivative corrections in 
 (\ref{ea44}). 
}
up to 2 loops
  \cite{FG,PSZ,JJN}: 
\be
|h|^2 \, \Big[ \hf \,(|q|^2 +  {1\over |q|^2}) 
-{1\over N^2}\, \Big|q  - {1\over q} \Big|^2 \Big] =g^2~.
\label{finiten1}
\ee
${}$For real $\beta$ deformation
or $|q|=1$, eq. (\ref{finiten1}) reduces to 
\be
|h|^2 \, \Big( 1 -{1\over N^2}\, \Big|q  - {1\over q} \Big|^2 \Big) =g^2~, 
\qquad \quad |q|=1~.
\label{finite2}
\ee
As was argued in  \cite{MPS}
using the analogy \cite{LM} with  the  non-commutative theory,  
in the {\it large $N$} limit,  the condition of finiteness  of the {\it real}
 deformation 
 (\ref{finite2}) or  $|h| =g$,  
is actually  the exact condition for conformal invariance to all loops. 
The condition of finiteness for complex deformation case  
  (\ref{finiten1}) is actually  true to three-loop order
\cite{FG,RSS} but is likely to  receive  higher-loop corrections.

Then it   follows from (\ref{deformed-masses}) that in the 
{\it real}  $\beta$-deformation case 
\be 
\cP \Big(  \cM_{(h,q)}^\dagger \cM_{(h,q)}   -|\cM_{(g,1)}  |^2 \Big)
~=~ O({1 \over  N})~,
\ee
where $\cP $ is  an  orthogonal projector on the subspace of charged 
states, 
\be 
\cP  \, E_{0\un{i}} =  E_{0\un{i}} ~, \qquad 
\cP  \, E_{\un{i} 0} =  E_{\un{i} 0} ~, \qquad 
\cP  \, E_{\un{i} \,\un{j} } = \cP\, H_I =0~.
\ee
One concludes  that in the {\it real} $\b$  case 
the deformation-dependent contribution coming from 
 the second line 
of (\ref{ea}) is subleading
 in the large $N$ limit:  one  is left then 
with  the contribution of the 
 first line of (\ref{ea}), which is just  the
 effective action in  the $\cN=4$ SYM case. 
 Equivalently,  the planar limit of the one-loop effective action in the 
 real deformation case does not depend on $\beta$.
 This will not be true in the  complex $\b$ case. 

Let us first    consider the case of  finite $N$. 
The effective action (\ref{ea}) 
is given  by the contributions from several $U(1)$-charged
superfields 
\bea
\G_{{\rm 1-loop}} &=& 
{\rm i} \,(N-1)\, {\bf tr }^{(e)}  \, 
\ln  \, ( \Box_{\rm v}
- |\m_{(g,1)} |^2)  \label{ea2} \\
&+& {\rm i} \,(N-1) \,{\bf tr}_+^{(e)}  
\ln  \, \Big[ (\Box_+ -  |\m_{(h,q)}^+|^2 )
\, (\Box_+ -  |\m_{(h,q)}^-|^2 )
( \Box_+   - |\m_{(g,1)} |^2)^{-2}  \Big]~, \non 
\eea
with $\m_{(h,q)}^\pm$ 
defined in (\ref{deformed-masses}).
The notation $ {\bf tr}^{(e)}$ (or ${\bf tr}_+^{(e)}$) 
indicates that the corresponding operator acts on the space
of  unconstrained (or chiral) superfields of charge $e$.

The transformations needed to put (\ref{ea2}) into an explicit 
proper-time representation form
are described in  Appendix C. 
The result is
\bea
\G_{{\rm 1-loop}} &=& 
-{\rm i} \,(N-1)
\int\limits_0^\infty  \frac{ {\rm d}  s }{  s} \,
\left(
 \int {\rm d}^8 z \,K(z,z|-{\rm i}s) \, {\rm e}^{ - s |\m_{(g,1)} |^2 }
\right. 
\label{ea3} \\
&&+ \left.
 \int {\rm d}^6 z \,K_+(z,z|-{\rm i}s) \Big[ 
 {\rm e}^{ -s |\m_{(h,q)}^+ |^2  } 
+  {\rm e}^{ -s |\m_{(h,q)}^- |^2   } 
-2\, {\rm e}^{ - s |\m_{(g,1)} |^2  } \Big]  \right)~,
\non
\eea
where we have Wick-rotated the proper-time integrals.
The background-dependent  heat kernels $K $ and $K_+$ 
are defined in Appendix C.  
As is seen  from (\ref{coincident}), 
there is no need to introduce a 
regularization -- the effective action is finite.
${}$ Following  
\cite{BKT}  and introducing the functions 
\bea
\o (x,y) &=& \o(y,x) \equiv 
\frac{\cosh x -1}{x^2} \, 
\frac{\cosh y -1}{y^2} \,
\frac{x^2 -y^2}{\cosh x - \cosh y} ~, \non \\
\z (x,y) &=& \z(y,x)\equiv 
-{1\over y^2} \Big\{ 
  \frac{\cosh x -1}{x^2} \, 
\frac{x^2 -y^2}{\cosh x - \cosh y} -1 \Big\}~,
\eea
one  finally obtains
\bea
 \G_{{\rm 1-loop}} &=&  
\frac{N}{16\p^2} 
 \ \int {\rm d}^6 z \,W^2  \,\ 
\ln \frac{  e^2}{ \s_{(h,q)}^+ \,\s_{(h,q)}^- } 
\non \\
&+&\frac{(N-1)}{8\p^2}  \int {\rm d}^8 z \,
\frac{  {\bar W}^2 W^2 }{ {\bar \f}^2 { \f } ^2}  
\int\limits_0^\infty   {\rm d}  s \,  s \, {\rm e}^{-s} \,
\Bigg[ \o (s \J /e, s{\bar \J}/e)  
\label{ea4} \\
&+& { \z (s \J / \s^+_{(h,q)}\, , 
s{\bar \J}/\s^+_{(h,q)}) \over 2 (\s^+_{(h,q)} /e)^2}
+ { \z (s \J / \s^-_{(h,q)} \,, 
s{\bar \J}/\s^+_{(h,q)}) \over 2 (\s^-_{(h,q)}/e)^2}
- { \z (s \J / e \,, 
s{\bar \J}/e ) 
} \Bigg]~.
\non 
\eea
Here we have  used the following notation:
\bea
{\bar \J}^2 = {1\over 4} D^2 \Big({W^2 \over {\bar \f}^2 { \f}^{2}} \Big)~, 
\qquad 
\J^2 ={1\over 4} {\bar D}^2 \Big({ {\bar W}^2\over  {\bar \f}^2 \f^{2}}
\Big)
~, 
\eea
and defined 
\be
\s^+_{(h,q)}\equiv 
e \, \Big|g^{-1} h \, \Big[
 q- {1\over N} \, (q  - {1\over q} )\Big] \Big|^2~, 
\qquad
\s^-_{(h,q)}\equiv 
e \, \Big|g^{-1} h \, \Big[
 {1\over q}+ {1\over N} \, (q  - {1\over q} )\Big] \Big|^2~.
\ee
The superfields $\J$ and $\bar \J$ are superconformal scalars
\cite{BKT} so that the functional (\ref{ea4}) is  invariant under the 
$\cN=1$ superconformal group. 
In (\ref{ea4}), $\f$ and $W_a$ 
may  no longer be assumed to obey the constant field approximation.

If the deformation parameter $\b$ is {\it real}, i.e.
 $|q|=1$, then $\s^+_{(h,q)}=\s^-_{(h,q)} \equiv \s_{(h,q)}$.
The condition of  conformal invariance (\ref{finite2}) implies 
  $\s_{(h,q)} =1 +O(1/N)$, so that, as already mentioned
  above, 
   in the large $N$ limit 
  the effective action for the real deformation reduces to that for
   the $\cN=4$ SYM theory, i.e. to  \cite{BKT}
\bea
&& \G_{{\rm 1-loop}} = 
\frac{N}{8\p^2}  \int {\rm d}^8 z \,
\frac{  {\bar W}^2 W^2 }{ {\bar \f}^2 \f^2}  
\int\limits_0^\infty   {\rm d}  s \,  s \, {\rm e}^{-s} \,
\Big[  \o (s \J , s{\bar \J}) ~+~O({1\over N} ) \Big]~.
\label{ea44} 
\eea

In the case of  general {\it complex} $\b$  deformation, i.e. 
$|q|\neq1$, there is no simple relationship 
between $ \G_{{\rm 1-loop}} $ and the effective action for $\cN=4$ SYM.
Keeping only the two- and four-derivative terms, 
the  finite $N$ effective action  takes the form
\bea
 \G_{{\rm 1-loop}}  & \approx &
\frac{N}{16\p^2}  \, C_2  \ 
 \int {\rm d}^6 z \,W^2
\ + \ 
\frac{(N-1)}{16\p^2} \, C_4 \   \int {\rm d}^8 z \,
\frac{  {\bar W}^2 W^2 }{ {\bar \f}^2 \f^2}  \ , 
\label{ea45} 
\eea
\bea
C_2= \ln \frac{ e^2}{ \s_{(h,q)}^+ \,\s_{(h,q)}^- } \ , \ \ \ \ \ \ \ 
C_4 =  1+ {1\over 12} \Big[
{e^2 \over  (\s^+_{(h,q)} )^2 }
+ {e^2 \over (\s^-_{(h,q)} )^2 } -2\Big] ~.
\eea
Using the finiteness  relation between $h,g$ and $q$ 
(\ref{finiten1}),  one finds  in the $N \to \infty$ limit  
\bea
\frac{ e^2}{ \s_{(h,q)}^+ \,\s_{(h,q)}^- } 
={1\over 4} \Big(|q|^2 +  {1\over |q|^2}\Big)^2  + O( { 1\over N}) ~, 
\eea
\bea
{e^2 \over  (\s^+_{(h,q)} )^2 }
+ {e^2 \over (\s^-_{(h,q)} )^2 } = 
{1\over 4} \Big(|q|^2 +  {1\over |q|^2}\Big)^4 
-\hf  \Big(|q|^2 +  {1\over |q|^2}\Big)^2 + O( { 1\over N})
~.
\eea
Thus  $ \G_{{\rm 1-loop}} $  depends on $|q|$ even in  the large $N$ limit.


\sect{Conclusions}
 
The above  computation  illustrates the difference between  the real and complex 
$\beta$ deformation 
cases. The real deformation   is obviously 
much closer to the $\N=4$ SYM theory.  Its  simplicity 
should have its origin in the  possibility  to give a  noncommutative theory 
interpretation to the real $\b$  deformation case \cite{LM},  
given that  the noncommutative theories are known to simplify in the large $N$ limit.

In particular, as discussed in the Introduction, the matching 
 between  the leading terms in the D3-brane probe action in the dual geometry of \cite{LM} and 
 in  the above 1-loop  large $N$  effective action for the real deformation case suggests 
 that the corresponding $F^2$ and $F^4$    non-renormalization theorems of undeformed 
 theory   continue to hold  in the large $N$ real deformation case, despite the  reduction in 
 the amount of supersymmetry from $\N=4$ to $\N=1$. 
 

It would obviously be interesting to generalize the above 
computation   of the effective action to more complicated 
 backgrounds 
for which the real $\beta$ deformation   dependence 
remains in  the large $N$ limit. 
This would allow one to probe 
the dual supergravity background of \ci{LM}  in a more  non-trivial 
way. 
In Appendix D we present the expression for 
the leading $W^2$ term in $\Gamma_{\rm 1-loop}$ in a more general diagonal background
that should correspond to several separated brane probes. 
 Another open problem is to  find the effective action in the
presence of the second exactly  marginal 
deformation $h' \sum_i \P^3_i $ of \ci{LS} for which the exact dual 
supergravity background is  not known at present. 
Another direction is generalization to deformations of
nonconformal theories, cf. \cite{LM,GN}.
\bigskip

\section*{Acknowledgments }
The work of S.M.K. is supported in part by the Australian 
Research Council.
The work of A.A.T.  was supported  by the DOE grant
 DE-FG02-91ER40690
and also by the INTAS contract 03-51-6346 and 
the RS Wolfson award.

\bigskip
\bigskip

\begin{appendix}

\sect{Background-field quantization}

To quantize the $\b$-deformed $\cN=4$ SYM theory
(\ref{b-deformed-N=4SYM}) we use the $\cN=1$ background 
field formulation \cite{GGRS}.
The first step is to 
implement the  background-quantum splitting 
(\ref{bq-splitting}).
Then the action becomes 
\bea
S= 
\int {\rm d}^8 z \, {\rm Tr} \Big(
( \F_i  +\vf_i{} )^\dagger \,
{\rm e}^{g\,v} \, ( \F_i +\vf_i) \, {\rm e}^{-g\, v} \Big)
&+&  {1\over g^2}\int {\rm d}^6 z \, {\rm Tr} \Big( \bW^\a \bW_\a  \Big) \non \\
+ \Big\{
\int {\rm d}^6 z \, \cL_{\rm c} (\F_i + \vf_i) 
&+& {\rm c.c.} \Big\}~,
\eea
where
$\cL_{\rm c}(\F_i)$ stands for the superpotential 
in   (\ref{b-deformed-N=4SYM}), and
\bea
\bW_\a &=& - {1\over 8} {\bar \cD}^2 \Big( 
{\rm e}^{-g\,v}\, \cD_\a \,{\rm e}^{g\,v} \cdot 1 \Big)  = \cW_\a
- {1\over 8} {\bar \cD}^2 \Big( 
g\, \cD_\a v- \hf \, g^2[v, \cD_\a v] 
\Big) + O(v^3)~.  
\eea 
Since both the gauge and matter background superfields are 
non-zero, 
it is convenient to use the 
$\cN=1$ supersymmetric 't Hooft gauge 
(a special case of the supersymmetric $R_\x$-gauge
introduced in \cite{OW} and further developed in \cite{BBP})
which is specified by the nonlocal gauge condition 
\bea
-4 \chi \,&= &{\bar \cD}^2 v +
g\, [\F_i, ({\Box_+})^{-1} {\bar \cD}^2 \vf_i{}^\dagger ] 
= {\bar \cD}^2 v +
g\,[\F_i, {\bar \cD}^2  ({\Box_-})^{-1}  \vf_i{}^\dagger ] 
~. 
\eea
Here  $\Box_+$ and  $\Box_-$ stand for
the covariantly chiral and antichiral d'Alembertians, 
\bea 
\Box_+ &=& \cD^a \cD_a - \cW^\a \cD_\a -\hf \, (\cD^\a \cW_\a)~, 
\quad
\Box_- = \cD^a \cD_a + {\bar \cW}_\ad {\bar \cD}^\ad 
+\hf \, ({\bar \cD}_\ad  {\bar \cW}^\ad)~.
\eea
The gauge conditions chosen lead 
to the following Faddeev-Popov ghost action 
\bea 
S_{\rm gh}
&=& {\rm Tr}
 \int {\rm d}^8 z \, (\tilde{c} -\tilde{c}^\dagger ) \left\{
L_{g v/2} \, (c+ c^\dagger ) 
+  L_{g v/2} \, \coth (  L_{gv/2} ) ( c- c^\dagger) 
\right\}   \\
&-& {\rm Tr}  \int {\rm d}^8 z \, \left\{ 
g^2 \,[\tilde{c}, \F_i] \, (\Box_-)^{-1}[c^\dagger , \F_i{}^\dagger 
+ \vf_i{}^\dagger]
+ g^2\,[ \tilde{c}^\dagger , \F_i{}^\dagger] \, (\Box_+)^{-1}
[c, \F_i + \vf_i ] \right\}~,
\non
\eea
with $L_X \, Y =[X,Y] $. 
Here the ghost (anti-commuting) superfields $c$ and  $\tilde{c}$  
are  background covariantly chiral.   One should add also 
the standard  gauge-fixing functional 
\be
S_{\rm gf} = - 
 \int {\rm d}^8 z \, {\rm Tr} \,(\chi^\dagger \,\chi)~, 
\ee
which is also accompanied by the 
   Nielsen-Kallosh ghost action 
\be
S_{\rm NK}
= \int {\rm d}^8 z \, {\rm Tr} \,(b^\dagger \,b) ~,
\label{third-ghost}
\ee
where the third (anti-commuting)  ghost  superfield $b$ 
is background-covariantly chiral.  
The Nielsen-Kallosh ghosts lead to  a  one-loop contribution only. 

\sect{Group-theoretical relations}

Let us  describe
the $SU(N)$ conventions adopted in this paper.
Lower-case Latin letters from the middle of the alphabet, 
$i,j,\dots$, 
are  used to denote the matrix elements in the fundamental
representation. 
We also set $i=(0,\un{ i})=0,1,\dots, N-1 $.
A generic element of the Lie algebra $su(N)$ is 
\be
u = u^I \, H_I + u^{ij} \, E_{ij} \equiv u^\m \,T_\m~,  
\qquad i \neq j ~. 
\label{generic}
\ee
We choose a Cartan-Weyl basis 
to consist of the elements:
\be 
H_I = \{ H_0, H_{\un{I}}\}~, \quad  \un{I} = 1,\dots, N-2~, 
\qquad \quad E_{ij}~, \quad i\neq j~. 
\label{C-W}
\ee 
The basis elements 
 defined as matrices in  the fundamental representation are   \cite{KM2}, 
\bea
(E_{ij})_{kl} &=& \d_{ik}\, \d_{jl}~, \non \\
(H_I)_{kl} &=& \frac{1}{\sqrt{(N-I)(N-I-1)} }
\Big\{ (N-I)\, \d_{kI} \, \d_{lI} - 
\sum\limits_{i=I}^{N-1} \d_{ki} \, \d_{li} \Big\} ~.
\label{C-W-2}
\eea
They satisfy 
\be
{\rm Tr} (H_I\,H_J) = \d_{IJ}\ , \ \ \ \ 
{\rm Tr} (E_{ij}\,E_{kl}) = \d_{il}\,\d_{jk}\ , \ \ \ \   
{\rm Tr} (H_I \,E_{kl}) =0 \ . \ee

\sect{Proper-time representation for the effective action} 

Here we  present  some technical details relevant for the 
evaluation of the effective action (\ref{ea2}); we follow 
 \cite{KM1}  where references to earlier work 
on covariant proper-time techniques in supersymmetric theories
can be found, see also \cite{BK}.

The effective action (\ref{ea2})
can be expressed in terms of  two different types of Green's 
functions in  a background  of a $U(1)$ vector multiplet 
 described
by the gauge covariant derivatives (\ref{gcd})
with
$\G_A = \G^\m_A (z)\, \bm{e}$.  
Here  $\bm e$ is a  charge operator, $\bm{e}= \pm e$. 
The corresponding gauge-invariant chiral field strength is
$\cW_\a = W_\a \, {\bm e}$.
Associated with $\Box_{\rm v} $  in   (\ref{ea2})  is the 
Green's function $G(z,z')$ 
\be
\Big(\Box_{\rm v} - |\m|^2 \Big) \, G(z,z') = - \d^8 (z-z')~,
\qquad 
G(z,z') = {\rm i} \int\limits_0^\infty {\rm d}s \, K(z,z'|s) \, 
{\rm e}^{ -{\rm i} |\m|^2 
s }~.
\label{proper-time-repr}
\ee
Associated with the  chiral d'Alembertian $\Box_+$ 
in   (\ref{ea2})
is the  Green's function $G_+(z,z'|s)$ which is covariantly 
chiral in both arguments, 
${\bar \cD}_\ad \, G_+(z,z') = {\bar \cD}'_\ad \, G_+(z,z') =0$; it 
satisfies 
\be
\Big(\Box_+ - |\m|^2 \Big) \, G_+(z,z') = - \d_+ (z,z')~, 
\qquad \d_+ (z,z') = -{1\over 4}{\bar \cD}^2\,\d^8(z-z')~.
\ee
This Green's function is generated by 
the chiral heat kernel $K_+(z,z'|s) $
which is introduced similarly to how this is done in  (\ref{proper-time-repr}).

The  heat kernel in (\ref{proper-time-repr})
has the following explicit form \cite{KM1} 
\bea
K(z,z'|s) &=& -\frac{\rm i}{(4 \pi s)^2} \, 
\sqrt{
\det
\left( \frac{2\, s \,\cF}{{\rm e}^{ 2  s \cF} -1}\right) } 
\; {U}(s) \,
\z^2  \bar{\z}^2 \,
{\rm e}^{ \frac{{\rm i}}{4} 
\r \, \cF \coth ( s \cF) \, \r } \, I(z,z')~.
\label{real-kernel}
\eea
Here 
\be 
{U}(s) = \exp \Big\{- {\rm i} s (\cW^{\a} \cD_{\a} 
+ \bar{\cW}^{\ad} {\bar \cD}_{\ad})\Big\}~,
\quad
I(z,z') = \exp \,\Big\{-{\rm i} 
\int_{z'}^{z} {\rm d}t \, \z^A \G_A (z(t)) \Big\}~, 
\ee 
where the integration is carried out along the straight line 
connecting the points $z'$ and $z$.  
The variables $\rho$ and $\zeta$  are components of the 
 supersymmetric two-point function 
$\z^A(z,z') 
\equiv (\r^a , \z^\a, {\bar \z}_\ad)$
 defined as 
\be
\r^a = (x-x')^a - {\rm i} (\q-\q') \s^a {\bar \q}' 
+ {\rm i} \q' \s^a ( {\bar \q} - {\bar \q}') ~, \ee
\be 
\z^\a = (\q - \q')^\a ~, \qquad
{\bar \z}_\ad =({\bar \q} -{\bar \q}' )_\ad ~. 
\label{two-point}
\ee
The chiral heat kernel 
$K_+(z,z'|s) $
is given by  \cite{KM1}
\bea
K_+(z,z'|s) &=& -{1 \over 4} {\bar \cD}^2 K(z,z'|s) =
-\frac{\rm i}{(4 \pi s)^2} \, 
\sqrt{ \det
\left( \frac{2\, s \,\cF}{{\rm e}^{ 2  s \cF} -1}\right) } 
\; {U}(s) \,
\non \\
& \times &
\z^2   \,
\exp 
\Big[ 
 \frac{{\rm i}}{4} \r \,  
\cF \coth ( s \cF) \, \r
-\frac{{\rm i}}{2} \r^a 
\cW   \s_a {\bar \z} \Big]
 \,  I(z,z')~.
\label{chiral-kernel}
\eea
For  the values of the heat kernels 
at coincident points  one obtains \cite{KM1}
\bea
-{\rm i}\,K(z,z|-{\rm i}s) 
&=& \frac{s^2}{(4 \pi )^2} \, \cW^2 \, {\bar \cW}^2\,
 {\sinh^2(s \cB/2) \over (s\cB/2)^2} \,
{\sinh^2(s {\bar \cB}/2) \over ( s{\bar \cB}/2)^2}\,
\sqrt{
\det
\left( \frac{s \,\cF}{\sin  (s \cF )}\right) }~,
 \non \\
-{\rm i}\, K_+(z,z|-{\rm i}s) &=& 
\frac{1}{(4\p)^2 }\,\cW^2\,
\frac{\sinh^2 (s \cB/2)}{(s \cB/2)^2} \,
\sqrt{ \det
\left( \frac{s \cF}{\sin(s \cF) } \right) } ~,
\label{coincident}
\eea
where we have introduced the notation 
\bea
\cB^2 = \hf {\rm tr}\, \cN^2~, \quad \cN_\a{}^\b = D_\a \cW^\b~,  
\qquad
{\bar \cB}^2 = \hf {\rm tr}\, {\bar \cN}^2~, 
\quad {\bar \cN}_\ad{}^\bd = {\bar D}_\ad {\bar \cW}^\bd~.
\eea
${}$For the background superfields under consideration, 
$\cB^2 = {1 \over 4} D^2 \cW^2$
and
${\bar \cB}^2 = {1 \over 4} {\bar D}^2 {\bar \cW}^2$.
One also finds that 
\be
\sqrt{ \det \left( \frac{s \,\cF}{\sin  (s \cF )}\right) }
=\hf \, \frac{s^2 (\cB^2 - {\bar \cB}^2)}
{ \cosh(s \cB) - \cosh(s {\bar \cB}) }~.
\ee
The resulting proper-time representation for the effective action is
then given by (\ref{ea3}).

\sect{Leading term in the effective action for a  more general 
diagonal background}

Here we present the expression for 
 the effective action for a  more general abelian 
 background than the one studied in section 3: we shall 
 allow the  diagonal entries  of the background fields to be different, i.e. 
\bea
\F = {1\over g}\,{\rm diag} \,(\f^1, \dots ,  \f^N) ~, 
\quad \cW_\a = {\rm diag} \,(W^1_\a , \dots ,W^N_\a) ~, 
\quad \sum_{i=1}^{N} \f^i =  \sum_{i=1}^{N} W^i_\a =0~.
\eea
The quantum superfields that couple to the background 
vector multiplet are associated with the off-diagonal generators
$E_{ij}$, 
\be
\cW_\a \,E_{ij} = (W^i_\a -W^j_\a) \,E_{ij}  
\equiv W^{[ij]}_\a \,E_{ij}~.
\ee
The mass operator (\ref{massop}) 
acts on the  generators  associated with  charged components as 
 \bea
\cM_{(h,q)} \, E_{ij } =   g^{-1} h \, \Big(
 q\, \f^i &-&  q^{-1} \, \f^j \Big) \, E_{ij}
\equiv \m^{[ij]}_{(h,q)} \, E_{ij} ~. 
\label{deformed-masses2}
\eea
The effective action (\ref{ea}) 
is given  by the sum of 
 contributions from several $U(1)$-charged
superfields 
\bea
\G_{{\rm 1-loop}} &=& {\rm i}
\sum_{i<j}  {\bf tr }^{[ij]}  \, 
\ln  \, ( \Box_{\rm v}
- |\m_{(g,1)}^{[ij]}  |^2)  \label{ea11} \\
&+&  {\rm i}  \sum_{i<j} {\bf tr}_+^{[ij]}  
\ln  \, \Big[ (\Box_+ -  |\m_{(h,q)}^{[ij]}|^2 )
\, (\Box_+ -  |\m_{(h,q)}^{[ji]}|^2 )
( \Box_+   - |\m_{(g,1)}^{[ij]} |^2)^{-2}  \Big]~, \non 
\eea
with $\m_{(h,q)}^{[ij]}$ 
defined in (\ref{deformed-masses2}).
The notation $ {\bf tr}^{[ij]}$ (or ${\bf tr}_+^{[ij]}$) 
indicates that the corresponding operator $\Box_{\rm v}$
(or $\Box_+ $) is assoicated with the $U(1)$ vector multiplet
of field strength $\cW_\a =  W^{[ij]}_\a$.
Eq. (\ref{ea11}) leads to the following 
proper-time representation:
\bea
\G_{{\rm 1-loop}} &=& 
-{\rm i} \sum_{i<j} 
\int\limits_0^\infty  \frac{ {\rm d}  s }{  s} \,
\left(
 \int {\rm d}^8 z \,K^{[ij]}(z,z|-{\rm i}s) \, 
{\rm e}^{ - s |\m_{(g,1)}^{[ij]} |^2 }
\right. 
\label{ea12} \\
&&\quad+ \left.
 \int {\rm d}^6 z \,K^{[ij]}_+(z,z|-{\rm i}s) \Big[ 
 {\rm e}^{ -s |\m_{(h,q)}^{[ij]} |^2  } 
+  {\rm e}^{ -s |\m_{(h,q)}^{[ji]} |^2   } 
-2\, {\rm e}^{ - s |\m_{(g,1)}^{[ij]} |^2  } \Big]  \right)~,
\non
\eea
Here the four- and higher-derivative
contributions can be written in a form similar 
to  the second and third lines in  (\ref{ea4}).
Compared to the first term   in (\ref{ea4}),  
the two-derivative part of (\ref{ea12}) has 
non-trivial dependence on the scalar field background 
\be
\G_{{\rm 1-loop}} \approx
 \frac{1}{16\p^2}  \,\int {\rm d}^6 z   \sum_{i<j}
\Big(W^{[ij]} \Big)^2 \,
\ln \Bigg[ \frac{ g^2  (   \f^i -   \f^j    )^2}{ h^2\, ( q\, \f^i -  q^{-1} \, \f^j )
( q^{-1}\, \f^i -  q \, \f^j )    } 
\Bigg]
~+~{\rm c.c.}~
\label{ea13}
\ee
A similar expression for the 1-loop effective action on the Coulomb 
branch  appeared   in  \cite{DH}.

${}$For the simplest  background (\ref{actual-background}), 
the right-hand side in (\ref{ea13}) reduces to 
the two-derivative part of (\ref{ea45}). Indeed, 
non-vanishing contributions occur only if
$i=0$ and $j =\un{j}$, and then  $W^{[0\un{j}]}_\a = e \,W_\a$, 
where $e =\sqrt{N /(N-1)}$. 

\end{appendix}

\small{

}


\begin{thebibliography}{99}

\bibitem{LM}
  O.~Lunin and J.~Maldacena,
  ``Deforming field theories with U(1) x U(1) global symmetry 
and their gravity  duals,''
  JHEP {\bf 0505}, 033 (2005)
  [hep-th/0502086].

\bibitem{LS}
  R.~G.~Leigh and M.~J.~Strassler,
  ``Exactly marginal operators and duality in four-dimensional N=1
  supersymmetric gauge theory,''
  Nucl.\ Phys.\ B {\bf 447}, 95 (1995) [hep-th/9503121].

\bi{old}
A.~Parkes and P.~C.~West,
  ``Finiteness in rigid supersymmetric theories,''
  Phys.\ Lett.\ B {\bf 138}, 99 (1984).
  ``Three-loop results in two-loop finite supersymmetric gauge theories,''
  Nucl.\ Phys.\ B {\bf 256}, 340 (1985).
D.~R.~T.~Jones and L.~Mezincescu,
  ``The chiral anomaly and a class of two-loop 
  finite supersymmetric gauge theories,''
  Phys.\ Lett.\ B {\bf 138}, 293 (1984).
D.~R.~T.~Jones and A.~J.~Parkes,
  ``Search for a three-loop finite chiral theory,''
  Phys.\ Lett.\ B {\bf 160}, 267 (1985).

\bi{ahar}
 O.~Aharony, B.~Kol and S.~Yankielowicz,
  ``On exactly marginal deformations of N = 4 SYM and type IIB  supergravity on
  AdS(5) x S5,''
  JHEP {\bf 0206}, 039 (2002)
  [hep-th/0205090].

\bibitem{NIPR}
  V.~Niarchos and N.~Prezas,
  ``BMN operators for N = 1 superconformal Yang-Mills theories and
    associated string backgrounds,''
  JHEP {\bf 0306}, 015 (2003)
  [hep-th/0212111].

\bibitem{FRT1}
  S.~A.~Frolov, R.~Roiban and A.~A.~Tseytlin,
  ``Gauge - string duality for superconformal deformations of N = 4 super
  Yang-Mills theory,''
  hep-th/0503192.
  ``Gauge-string duality for (non)supersymmetric deformations 
of N = 4 super  Yang-Mills theory,''
 hep-th/0507021.
S.~Frolov,
  ``Lax pair for strings in Lunin-Maldacena background,''
  JHEP {\bf 0505}, 069 (2005)
  [hep-th/0503201].


\bi{mat}
T.~Mateos,
  ``Marginal deformation of N = 4 SYM and Penrose limits with continuum
  spectrum,''
  hep-th/0505243.
R.~de Mello Koch, J.~Murugan, J.~Smolic and M.~Smolic,
  ``Deformed PP-waves from the Lunin-Maldacena background,''
  hep-th/0505227.

\bibitem{RR}
  R.~Roiban,
  ``On spin chains and field theories,''
  JHEP {\bf 0409}, 023 (2004)
  [hep-th/0312218].
  D.~Berenstein and S.~A.~Cherkis,
  ``Deformations of N = 4 SYM and integrable spin chain models,''
  Nucl.\ Phys.\ B {\bf 702}, 49 (2004)
  [hep-th/0405215].
N.~Beisert and R.~Roiban,
  ``Beauty and the twist: The Bethe ansatz for twisted N = 4 SYM,''
  hep-th/0505187.

  
  
\bibitem{FG}
  D.~Z.~Freedman and U.~Gursoy,
  ``Comments on the beta-deformed N = 4 SYM theory,''
  hep-th/0506128.

\bibitem{PSZ}
  S.~Penati, A.~Santambrogio and D.~Zanon,
  ``Two-point correlators in the beta-deformed N = 4 SYM 
at the next-to-leading  order,''
  hep-th/0506150.

\bibitem{RSS}
  G.~C.~Rossi, E.~Sokatchev and Y.~S.~Stanev,
  ``New results in the deformed N = 4 SYM theory,''
  hep-th/0507113.

\bibitem{MPS}
  A.~Mauri, S.~Penati, A.~Santambrogio and D.~Zanon,
  ``Exact results in planar N=1 superconformal Yang-Mills theory,''
  hep-th/0507282.

\bibitem{BE}
  D.~Berenstein, V.~Jejjala and R.~G.~Leigh,
  ``Marginal and relevant deformations of N = 4 field theories and
  non-commutative moduli spaces of vacua,''
  Nucl.\ Phys.\ B {\bf 589}, 196 (2000)
  [hep-th/0005087].

\bibitem{DH}
  N.~Dorey and T.~J.~Hollowood,
  ``On the Coulomb branch of a marginal deformation of N = 4 SUSY Yang-Mills,''
  JHEP {\bf 0506}, 036 (2005)
  [hep-th/0411163].

\bi{mald}
J.~M.~Maldacena,
  ``Branes probing black holes,''
  Nucl.\ Phys.\ Proc.\ Suppl.\  {\bf 68}, 17 (1998)
  [hep-th/9709099].
  ``Probing near extremal black holes with D-branes,''
  Phys.\ Rev.\ D {\bf 57}, 3736 (1998)
  [hep-th/9705053].
I.~Chepelev and A.~A.~Tseytlin,
  ``Interactions of type IIB D-branes from the D-instanton matrix model,''
  Nucl.\ Phys.\ B {\bf 511}, 629 (1998)
  [hep-th/9705120].


\bibitem{BKT}
  I.~L.~Buchbinder, S.~M.~Kuzenko and A.~A.~Tseytlin,
  ``On low-energy effective actions in N = 2, 4 superconformal 
theories in  four dimensions,''
  Phys.\ Rev.\ D {\bf 62}, 045001 (2000)
  [hep-th/9911221].
I.~L.~Buchbinder, A.~Y.~Petrov and A.A.~Tseytlin,
  ``Two-loop N = 4 super Yang Mills effective 
  action and interaction  between
  D3-branes,''
  Nucl.\ Phys.\ B {\bf 621}, 179 (2002)
  [hep-th/0110173].

\bi{DS}
M.~Dine and N.~Seiberg,
  ``Comments on higher derivative operators in some SUSY field theories,''
  Phys.\ Lett.\ B {\bf 409}, 239 (1997)
  [hep-th/9705057].

\bibitem{BK} I.~L.~Buchbinder and S.~M.~Kuzenko,
{\it Ideas and Methods of Supersymmetry and
Supergravity or a Walk Through Superspace},
IOP, Bristol, 1998.

\bibitem{GGRS}
S.~J.~Gates, M.~T.~Grisaru, M.~Ro\v{c}ek and W.~Siegel,
{\it Superspace, Or One Thousand 
and One Lessons in Supersymmetry},
Benjamin/Cummings, 1983 [hep-th/0108200].

\bibitem{KM2}
  S.~M.~Kuzenko and I.~N.~McArthur,
  ``On the two-loop four-derivative quantum corrections in 4D N = 2
  superconformal field theories,''
  Nucl.\ Phys.\ B {\bf 683}, 3 (2004)
  [hep-th/0310025];
 ``Relaxed super self-duality and N = 4 SYM at two loops,''
  Nucl.\ Phys.\ B {\bf 697}, 89 (2004)
  [hep-th/0403240].

\bibitem{OW}
B.~A.~Ovrut and J.~Wess,
``Supersymmetric $R_\x$ gauge and radiative symmetry breaking,''
Phys.\ Rev.\ D {\bf 25} (1982) 409;
P.~Binetruy, P.~Sorba and R.~Stora,
``Supersymmetric S covariant $R_\x$ gauge,''
Phys.\ Lett.\ B {\bf 129} (1983) 85.

\bibitem{BBP}
A.~T.~Banin, I.~L.~Buchbinder and N.~G.~Pletnev,
``Low-energy effective action in N = 2 super 
Yang-Mills theories on  non-abelian background,''
Phys.\ Rev.\ D {\bf 66}, 045021 (2002) 
[hep-th/0205034];
``One-loop effective action for N = 4 SYM theory 
in the hypermultiplet  sector: Leading low-energy 
approximation and beyond,''
Phys.\ Rev.\ D {\bf 68}, 065024 (2003) 
[hep-th/0304046].

\bibitem{JJN}
  I.~Jack, D.~R.~T.~Jones and C.~G.~North,
  ``$N=1$ supersymmetry and the three-loop anomalous dimension 
  for the chiral superfield,''
  Nucl.\ Phys.\ B {\bf 473}, 308 (1996)
  [hep-ph/9603386].

\bibitem{KM1}
  S.~M.~Kuzenko and I.~N.~McArthur,
  ``On the background field method beyond one loop: 
A manifestly covariant
  derivative expansion in super Yang-Mills theories,''
  JHEP {\bf 0305}, 015 (2003)
  [hep-th/0302205];
  ``Low-energy dynamics in N = 2 super QED: Two-loop approximation,''
  JHEP {\bf 0310}, 029 (2003)
  [hep-th/0308136].


\bibitem{GN}
  U.~Gursoy and C.~Nunez,
  ``Dipole deformations of N = 1 SYM and supergravity backgrounds 
with U(1) x  U(1) global symmetry,''
 hep-th/0505100.


\end{thebibliography}
\end{document}